# Broadband ptychography using curved wavefront illumination

DANIEL S. PENAGOS MOLINA,[1,2,3,*] LARS LOETGERING,[1,4]
WILHELM ESCHEN,[1,2,3] JENS LIMPERT,[1,2,3,5] JAN ROTHHARDT[1,2,3,5]

[1]*Institute of Applied Physics and Abbe Center of Photonics, Friedrich-Schiller-University Jena,Albert-Einstein-Straße 15, 07745 Jena, Germany*
[2]*Helmholtz-Institute Jena, Fröbelstieg 3, 07743 Jena, Germany*
[3]*GSI Helmholtzzentrum für Schwerionenforschung, Planckstraße 1, 64291 Darmstadt, Germany*
[4]*Carl Zeiss AG, Carl Zeiss Promenade 10, 07745 Jena, Germany*
[5]*Fraunhofer Institute for Applied Optics and Precision Engineering, Albert-Einstein-Straße 7, 07745 Jena, Germany*
[*]*santiago.penagos@uni-jena.de*

**Abstract:** We examine the interplay between spectral bandwidth and illumination curvature in ptychography. By tailoring the divergence of the illumination, broader spectral bandwidths can be tolerated without requiring algorithmic modifications to the forward model. In particular, a strong wavefront curvature transitions a far-field diffraction geometry to an effectively near-field one, which is lees affected by temporal coherence effects. The relaxed temporal coherence requirements allow for leveraging wider spectral bandwidths and larger illumination spots. Our findings open up new avenues towards utilizing pink and broadband beams for increased flux and throughput at both synchrotron facilities and lab-scale beamlines.

## 1. Introduction

Ptychography [1, 2] is a scanning version of coherent diffractive imaging (CDI), in which the object is illuminated in overlapping positions by a confined illuminating beam. It has found applications in a wide spectral range, including terahertz [3], near-infrared (NIR) [4], visible spectrum (VIS) [5], extreme ultraviolet (XUV) [6], soft X-ray [7], and hard X-rays [8, 9]. While single-shot CDI requires a highly coherent illumination [10], ptychography can be used to robustly account for and mitigate decoherence effects [11–13]. For both single-shot CDI and ptychography, algorithmic and experimental techniques have been developed to account for reduced temporal coherence. These include physical modeling of dispersion effects in free-space propagation [14–19], tailoring or accounting for wavelength-dependent wavefronts to retrieve multiple spectral components in the illumination [13, 20–22], or using techniques from Fourier transform spectroscopy [23–25]. While the aforementioned techniques allow for utilizing broadband radiation in ptychographic beamlines, they come at the cost of either algorithmic overhead or interferometric hardware modifications. Consequently, the question arises under which conditions a polychromatic forward model or hardware modifications are needed at all. Early work on single-shot CDI suggested a trade-off between spectral bandwidth and spatial resolution [26]. This criterion has been utilized for ptychography [27] as a rule-of-thumb. Furthermore, recent work [28] has empirically demonstrated that ptychography exhibits a higher tolerance for broadband conditions in the near-field as compared to far-field (Fraunhofer) diffraction geometries. However, switching to a near-field geometry in the short-wavelength regime such as XUV or x-rays is challenging due to physical constraints that limit the minimum object to detector distance. In our work, we explore how beam curvature can serve as an additional degree of freedom that relaxes coherence requirements, thereby allowing for larger spectral bandwidths and potentially faster scanning and higher throughput in ptychography experiments utilizing sources with low temporal coherence.

The purpose of this work is threefold: First, We examine the spectral bandwidth criterion for single-shot CDI and investigate its validity for ptychography under varied experimental conditions and the assumption of a monochromatic forward model. Second, We continuously transition between near- and far-field diffraction geometries by controlling the illumination curvature – a mechanism that can be harnessed in short-wavelength experimental setups without any hardware modifications. Finally, We examine implications for dispersive specimens. The latter point is critical: does tolerance to broad bandwidth conditions imply insensitivity to dispersion effects? We demonstrate that tolerance to broad bandwidth conditions and information retrieval about dispersion are not mutually conflicting goals. Our findings are supported by both simulation and experiment.

The paper is structured as follows: In Section 2, we revisit the geometrical arguments that lead to the single-shot CDI spectral bandwidth versus spatial resolution restriction, and provide an alternative theoretical viewpoint, where we discuss the stationary phase approximation, which provides a mechanism to understand how temporal coherence requirements are relaxed under increasing illumination curvature. In Section 3, we present experimental results that demonstrate improved tolerance to broadband conditions under curved illumination conditions. The results are generalized by a simulation study on dispersive specimens in Section 4. Our findings are discussed and summarized in Section 5.

## 2. Broadband diffraction considerations

### 2.1. Geometrical considerations

In this section we revisit an inequality between the bandwidth and the achievable lateral resolution in a single-shot CDI setup, originally described in [26]. With regard to Fig. 1, consider two laterally displaced point sources on a specimen. When sub-waves originating from these two points interfere on a distant detector, the contrast tends to vanish as the optical path length is increased. In fact, writing down an explicit expression for the optical path length difference and using the paraxial approximation the authors of [26] estimated that when the temporal coherence length is larger than the optical path length difference, the following inequality holds true:

$$\frac{\Delta\lambda}{\lambda_c} < \frac{\delta}{D}. \tag{1}$$

Here $\delta$ is the smallest resolvable feature, $D$ is the lateral size of the illumination, and $\Delta\lambda$ and $\lambda_c$ are the spectrum's bandwidth and its center wavelength respectively. Inequality (1) predicts a linear behavior between the spectral bandwidth and the smallest resolvable feature, namely that we should expect a twofold loss in spatial resolution if we double the bandwidth in a single-shot CDI experiment.

Since the smallest resolvable feature $\delta$ is bounded by the wavelength and numerical aperture of the setup geometry, Eq. 1 can be written in terms of the numerical aperture *NA* of the system,

$$D \cdot NA < \frac{\lambda_c^2}{\Delta\lambda}, \tag{2}$$

where the right term of the inequality $\lambda_c^2/\Delta\lambda$ is an estimation of the temporal coherence length of the source [29].

### 2.2. Stationary phase approximation

In this section, we analyse the oscillatory behavior of the Fresnel diffraction integral under curved wavefront illumination and apply the method of stationary phase to explain its implications for temporal coherence requirements. Consider a two-dimensional electric field $E$ diffracted from an aperture (see Fig. 1). Under the paraxial approximation, the diffracted field at a distance $z$ can be expressed in the form of the following Fresnel diffraction integral [30]

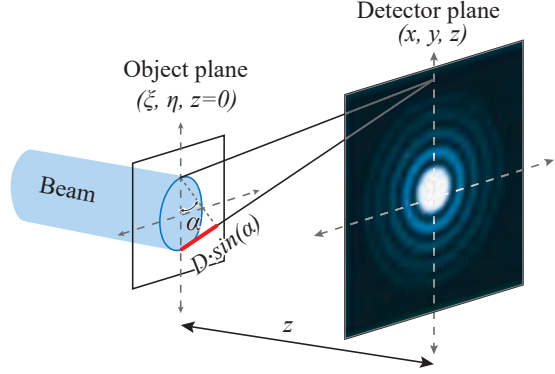

Fig. 1. The schematic illustrates the largest optical path difference $\text{OPD}_{\text{max}} \approx D \cdot sin(\alpha)$ from the outermost points in an illuminating beam with lateral size $D$ towards the outermost pixel in the detector. The contrast of the fringes produced by these points degrades as the OPD exceeds the temporal coherence of the source.

$$E(x, y, z) = \frac{e^{ikz}}{i\lambda z} \iint\limits_{\xi, \eta} \psi(\xi, \eta, 0) \exp\left[\frac{ik}{2z}\left[(x-\xi)^2 + (y-\eta)^2\right]\right] d\xi d\eta, \quad (3)$$

where $(\xi, \eta)$ and $(x, y)$ are the coordinates in the object and the observation plane, respectively. Evaluating this integral in regions where $|\xi| \gg |x|$ and $|\eta| \gg |y|$ leads to rapid phase oscillations between $-\pi$ and $\pi$ that cancel each other [30,31] (see Fig. 2). The key element of the stationary phase approximation (SPA) is to evaluate the integral in Eq. 3 in an effectively smaller domain where the phase is slowly oscillating. Further insight into Eq. 3 is gained when taking into account an explicit expression for the illumination, which is implicit in the exit field $E$ in the integrand of Eq. 3. Here we assume a quadratic phase illumination, which can model convergent and divergent behaviour of the illumination, we have

$$\psi(\xi, \eta) = P(\xi, \eta) \, O(\xi, \eta) \exp\left[\frac{ik}{2f}[\xi^2 + \eta^2]\right], \quad (4)$$

where $O(\xi, \eta)$ denotes the complex object transmission, $P(\xi, \eta)$ the illumination amplitude and $\exp\left[\frac{ik}{2f}[\xi^2 + \eta^2]\right]$ its curvature as a quadratic phase exponential. In this form it can be seen that it modifies the effective curvature of the Fresnel integration kernel in Eq. 3. The parameter $f$ is the distance to a focal spot up- or downstream of the object. When the beam is focused at the detector plane (i.e, $f = -z$), Eq. 4 cancels out the quadratic phase term in a near-field geometry and make the diffraction appear as a far-field one [30]. Conversely, the quadratic phase of the illuminating beam can convert a far-field diffraction pattern to effectively resemble a near-field diffraction pattern. Evaluating Eq. 3 with the phase term in Eq. 4 reveals that the asymptotic behavior can be reached faster (see Fig. 2), when a strongly curved illumination wavefront is used. Effectively, this implies a decrease of the integration region with increasing beam curvature, as illustrated in Fig. 2. This has two important bearings for ptychography: (1) for monochromatic beams, beam curvature increases the spatial frequency content in the illumination and relaxes the dynamic range requirements of the detector, and increases the diffraction-limited resolution [32, 33]; (2) for polychromatic beams, according to the SPA beam curvature shrinks the domain from which different parts of the beam effectively interfere. The latter stands in conflict with the simple relation discussed in the last section (Eq. 2), which assumes a fixed beam diameter. The SPA on the other hand predicts an effectively shrinking beam diameter, thus resulting in relaxed coherence requirements due to decreased optical path differences. In what follows we experimentally validate the relaxation of temporal coherence requirements for ptychography, as predicated by the SPA.

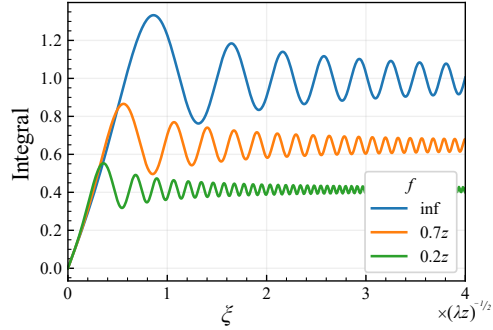

Fig. 2. One-dimensional evaluation of Eq. 3 under the influence of an additional beam curvature (Eq. 4). As the distance $f$ to a focal spot upstream the object increases (i.e, divergence of beam decreases), the integral reaches its asymptotic value for a smaller integration range $\xi$.

## 3. Experiments

### 3.1. Setup

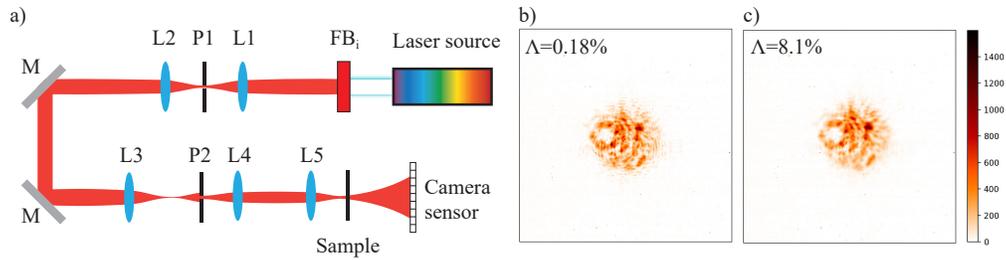

Fig. 3. a) Scheme of the experimental setup. A white light source is filtered by means of a selectable bandpass filter FB$_i$ in order to vary the spectral bandwidth of the source. Subsequently the beam is spatially filtered and expanded (L1-P1-L2). A second pinhole P2 is relayed onto the sample with a 4f configuration (L4-L5). The key role is played by L3, whose position determines the curvature of the wavefront incident on P2 and thus controls the divergence of the illumination on the sample. b-c) Raw data for beam curvature $f$ =0.9 cm measured with $\Lambda = 0.18\%$, and $\Lambda = 8.1\%$, respectively.

The experimental setup is shown in Fig. 3a). A super-continuum laser source was spectrally filtered by means of bandpass filters, resulting in four different center wavelengths ($\lambda_c$ =[632.8 nm, 650 nm, 643.95 nm, 636.57 nm]) and 4 different relative bandwidth ($\Lambda = \Delta\lambda/\lambda_c$) conditions (0.18%, 0.95%, 3.3%, 8.1%). The beam was spatially filtered (L1-P1-L2) to ensure effects due to reduced spatial coherence can be ignored. To this end P1 was set to $\approx$30% larger than the diffraction-limited resolution of L1. A 500 µm pinhole (P2) was imaged onto the specimen using a 4f-system. The detector downstream of the object had a pixel-size of 3.45 µm × 3.45 µm, an array size of 4438 × 4438, and a 12-bit dynamic range. The divergence of the illumination was controlled by a movable lens (L3) upstream of the pinhole (P2), which resulted in an illumination

$NA_{illu}$=[0.005, 0.02, 0.027] for its corresponding curvature $f_i$=[5 cm, 1.3 cm, 0.9 cm]. An example of the raw data for $f$ =0.9 cm showing the spectral blurring due to the increased spectral bandwidth is shown in Fig. 3b-c). Four different bandwidth conditions and three different phase curvatures were used (Fig. 4[a-c]). The sample-detector distance was found to be 55.28 mm using the zPIE axial position calibration algorithm [34], which resulted in an NA≈0.14 and a lateral resolution limit of 2.3 µm. To generate comparable photon flux conditions between all 12 combinations for curvatures and bandwidths, the total number of intensity counts for each dataset was held constant. This resulted in a lower signal to noise ratio for the case of the more divergent beam. Each dataset consisted of 100 scan positions arranged in a Fermat spiral [35] with an average linear overlap [36] of 90% between consecutive positions. A USAF resolution test target was chosen as a specimen to allow for comparing the lateral resolution between each reconstruction.

### 3.2. Results

The open-source software *PtyLab* [37] was used to run 300 iterations of mPIE [38] using the same initial parameters for the reconstruction of all datasets. The reconstruction results are shown in Fig. 4. For all reconstructions an inset of the same region was inserted for comparison. Two general tendencies are seen from the reconstructed micrographs: First, the lateral resolution decreases with increasing bandwidth (left to right columns). Second, the loss of lateral resolution can be compensated by increasing the wavefront curvature (bottom to top row). Thus our main finding is that a larger spectral bandwidth can be tolerated when the curvature of the illumination increases. Comparing, for instance, conditions a-$\Lambda_3$ with c-$\Lambda_2$, we see that a comparable imaging resolution is achieved, albeit using a 3 times larger relative bandwidth in the former. To evaluate the achieved resolution, we applied the Rayleigh criterion to a lineout taken across the smallest features of the reconstructed USAF target, specifically where the visibility of the features exceeded 20%. Next, we identified the group and number of these features and computed the full pitch resolution using the USAF target formula. An overview of the lateral resolution $\delta$ for each panel is summarized in Table 1 and Fig. 5. The columns and rows in Table. 1 are organized in the same manner as in Fig. 4. For easy comparison, a relative resolution $\delta/D$ represented in percentage was added. The blue shaded cells mark where Eq. 1 holds, while the red cells indicate that this inequality was violated. Thus we find both circumstances where Eq. 1 is outperformed and other situations where the lower resolution bound is not reached.

| | Curvature | $\Lambda_1$ = 0.18 % | | $\Lambda_2$ = 0.95 % | | $\Lambda_3$ = 3.3 % | | $\Lambda_4$ = 8.1 % | |
|---|---|---|---|---|---|---|---|---|---|
| | | $\delta$ | $\delta/D$ | $\delta$ | $\delta/D$ | $\delta$ | $\delta/D$ | $\delta$ | $\delta/D$ |
| **A** | $f \approx 0.9$ cm | 11.0 µm | 2.2 % | 11.0 µm | 2.2 % | 15.6 µm | 3.1 % | 19.7 µm | 3.9 % |
| **B** | $f \approx 1.3$ cm | 11.0 µm | 2.2 % | 12.4 µm | 2.5 % | 17.5 µm | 3.5 % | 22.1 µm | 4.4 % |
| **C** | $f \approx 5.0$ cm | 12.4 µm | 2.5 % | 14.0 µm | 2.8 % | 19.7 µm | 3.9 % | 24.8 µm | 5.0 % |

Table 1. Comparison of spatial resolution $\delta$ for each relative bandwidth $\Lambda_i$ and a given beam diameter $D \approx 500$ µm. The color of the cells indicate whether the inequality in Eq. 1 is either respected (*blue*) or violated (*red*). A line-plot of this table is shown in Fig. 5.

## 4. Dispersive sample considerations

The last section showed that a beam with a curved wavefront can increase the tolerance to a broader spectrum, provided that we assume a monochromatic forward model. However, ptychography is well-known to allow for extracting polychromatic wavefronts as well as individual color channels of dispersive specimens [13, 18, 20–22]. Thus we must ask the question how curvature affects our ability to extract individual spectral components if we assume a polychromatic forward model. Does increasing wavefront curvature entail a loss of sensitivity to polychromatic information? While in the last section the specimen was non-dispersive and the wavefront for individual spectral components is not expected to change significantly, the situation would certainly be different if the specimen was dispersive. Under such circumstances we would want to be able to spectrally resolve the specimen [13, 39] and extract the individual spectral wavefronts [18, 21, 22]. To this end, we simulated a ptychographic experiment using a dispersive sample and a trichromatic, discrete spectrum.

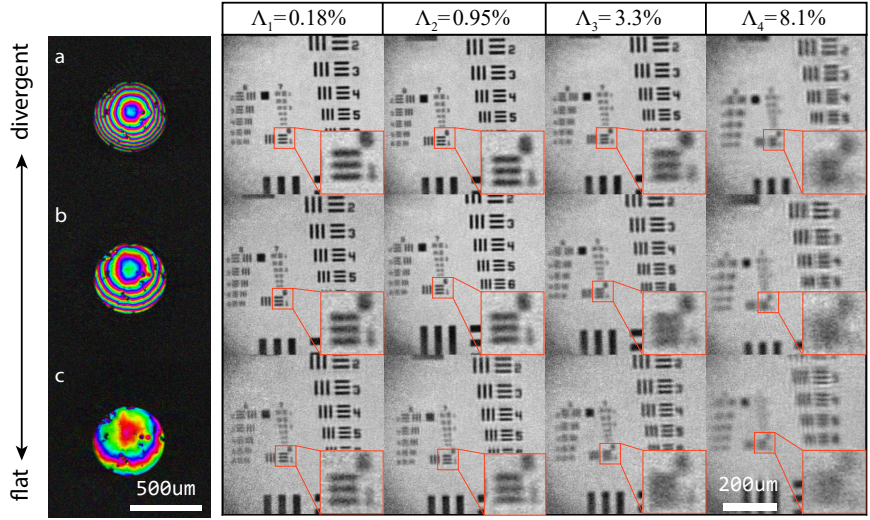

Fig. 4. Comparison of ptychographic reconstructions using 3 different wavefront curvatures, each for 4 different relative bandwidths $\Lambda_i = \Delta\lambda / \lambda_c$. *Left column*, reconstructed beam profile (a-c) for variable illumination curvatures. Two general tendencies are seen from the reconstructed micrographs: (1) The lateral resolution decreases with increasing bandwidth (left to right columns). (2) The loss of lateral resolution can be compensated by increasing the wavefront curvature (bottom to top row).

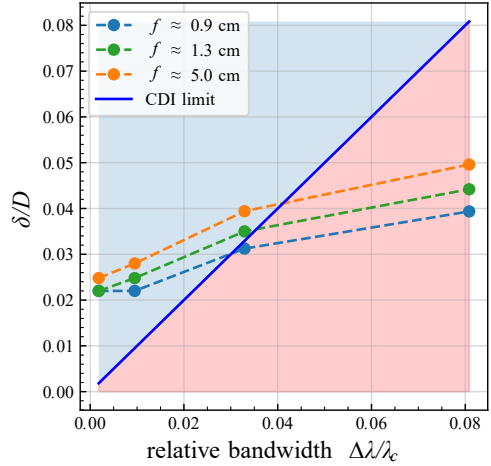

Fig. 5. Lineplot results from the bandwidth vs curvature experiments. The theoretical CDI limit (i.e. Eq. 1) predicts all points to be in the blue-shaded region. Which is not the case for our experimental data. As stated before, a stronger curvature, i.e, a small $f$, seems to increase the tolerance to larger spectral bandwidths and the spatial resolution improves.

The simulated beam consisted of three spectral lines at wavelengths 600 nm, 650 nm, and 700 nm. For the dispersive sample, we used a Siemens star resolution test target with the refractive index values of NBK7 glass and a binary thickness profile of 632 nm that results in a phase delay of OPD ≈ π radians at λ = 650 nm. We also included an amplitude spectral response from this sample with a custom coating layer, whose transmissivity increases towards longer wavelengths such that we expect 50% transmission at the central wavelength, as shown in Fig. 6a-c). With the aforementioned parameters we expected a distinct object spectral response (amplitude and phase shift) for each spectral line, as depicted in Table. 2. The setup parameters were chosen to be similar to our experimental setup (i.e, beam diameter $D$=500 μm, distance to detector $z$=55.28 mm, detector size $L$=12.8 mm, 200 scans positions with an average linear overlap of ∼ 80% between consecutive positions).

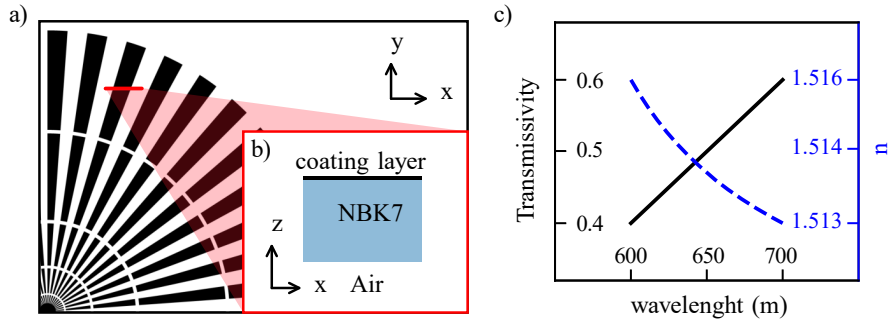

Fig. 6. a) Scheme and cross-section profile b) of the simulated Siemens-star target. c) Refractive index and transmissivity used for the substrate and the coating layer.

|                | **600 nm** | **650 nm** | **700 nm** |
|----------------|------------|------------|------------|
| Transmissivity | 0.4        | 0.5        | 0.6        |
| OPD (rad)      | 3.42       | 3.14       | 2.91       |

Table 2. Expected spectral response of dispersive sample for the simulated comb spectrum.

The simulations were carried twice, once with a curved beam ($f$=20 mm), and then with a flatter illumination ($f$=100 mm). The reconstructed results for both of these conditions are shown in Fig. 7. On the columns (a-c, g-i) we see the object spectral response, where the color represent the phase shift and amplitude is encoded in the brightness. Next to each object is the beam profile for each spectral line plotted in the same color scale as the object. At first sight there is no clear difference between the curved and flat beam object reconstructions. However, to assess the differences closely, we analyzed the histogram of the reconstructed object's amplitude and phase from each spectral line (see Fig. 8a), and fitted a Gaussian mixture model to evaluate the shape and location of the reconstructed amplitude and phase values against the expected values in Table. 2. We found that the histogram's peaks of both reconstructions closely match the ground truth spectral response (black triangles). However, the histogram from the data set with a flatter beam (*bottom row*) shows wider lobes in both the amplitude- and phase-distributions, indicating a larger uncertainty in the reconstruction for the flat beam. On the contrary, the histograms retrieved with a curved beam (*top row*) show no significant artifacts, which is important for quantitative results and material specific imaging [40]. In order to exclude spurious oscillations from difficult to resolve high spatial frequencies, these histograms were extracted from a region of interest mainly containing mid-spatial frequencies, as shown in Fig. 8b. Thus the main finding from this section is that curved beams do not entail a loss of sensitivity to polychromatic information. As in the monochromatic case, they help improve the reconstruction quality also for multi-spectral wavefronts.

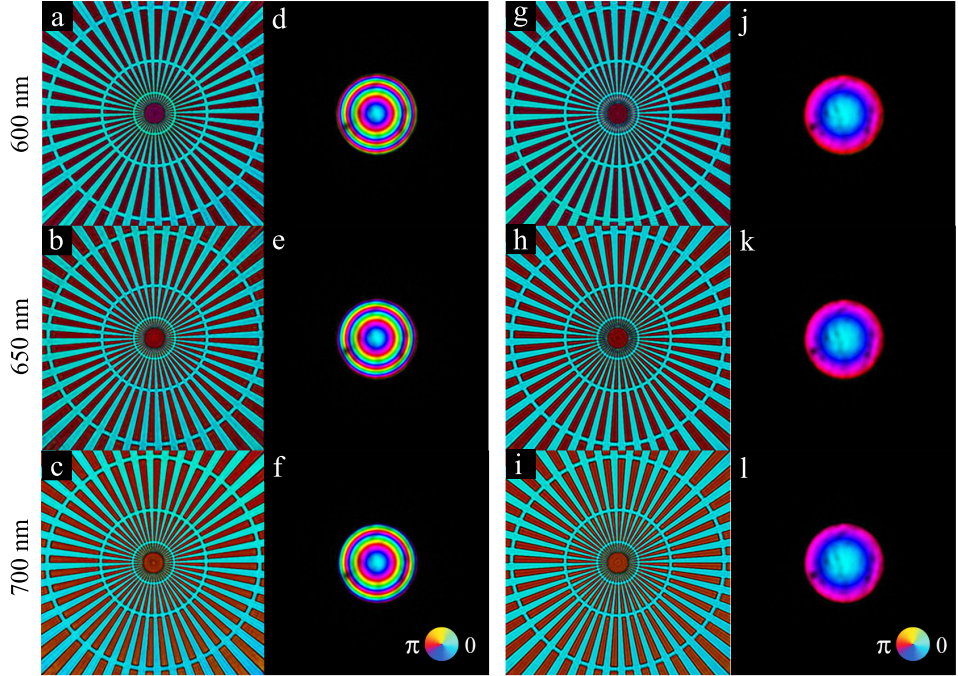

Fig. 7. Polychromatic reconstructions of the dispersive sample for a curved beam with $f = 20$ mm (a-f) and flatter one with $f = 100$ mm (g-l). Each row corresponds to a specific wavelength. At first glance, there is no clear qualitative difference on the reconstructed object between corresponding colors.

It should be noticed that for the experimental results were obtained with a broad continuous spectrum while the numerical simulations presented here were performed with three discrete wavelengths – which is a simplification of a typical spectrum provided with HHG sources [20, 22]. By refraining from simulating continuum spectra we circumvent another open question in ptychography, namely: how many discrete spectral points are required to represent a continuum spectrum? Previous work has already shown the capabilities to use ptychography with such spectra and reconstruct it a certain number of colors [18, 19, 41]. However, the criteria for choosing n-number of wavelengths to represent the input spectrum is still not clear. Thus, we restricted the simulations to discrete colors, such that the reconstructed object spectral response was not influenced by the aforementioned interrogation.

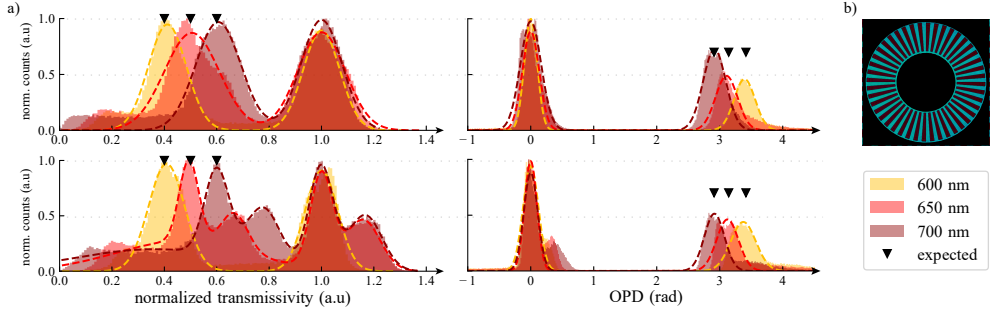

Fig. 8. a) Amplitude and phase histograms obtained from the polychromatic reconstructions. *Top row* Histograms for $f = 20$ mm (high curvature). *Bottom row* Histograms for $f = 100$ mm (low curvature). b) Selected region from which the histograms where computed. The black markers show the expected spectral response according to Table. 2. A cleaner histogram, was obtained using a smaller $f$.

## 5. Discussion and conclusion

In this work, we expanded upon the experimental observations of [28], and observed an improved tolerance of ptychography to a broad spectrum under a monochromatic forward model for increasingly divergent beams. Furthermore, we found that the commonly used temporal coherence limit for single shot CDI [26] can be outperformed in ptychography, which is consistent with previous work [18, 27]. This is clear evidence that the spectral bandwidths constraints in ptychography are more relaxed than in single shot CDI, and that curved beams affect this tolerance. However, we are unable to provide a theoretical bound for a rigorous spectral bandwidth constraint that holds for ptychography. Further investigations in this direction will be required in future work.

We presented an intuitive explanation for the increased tolerance to broadband illumination under increasing wavefront curvature via the stationary phase approximation. In particular, the effective region of integration in the Fresnel diffraction integral shrinks as curvature increases. This translates into an effectively smaller beam, which explains why the simplified single shot CDI rule of thumb (Eq. 2, which assumes a fixed beam diameter), is not valid. In other words, larger optical path differences do not contribute to the integral and the temporal coherence requirement is relaxed.

We extended our investigation to polychromatic specimens and wavefronts. A naive extrapolation of the previous finding could suggest that curved beams are harmful for polychromatic experiments, as they would inhibit sensitivity to dispersion effects under a monochromatic forward model. However, our simulations indicate that that beam curvature does not result in a loss of spectral sensitivity, in spite of the previously mentioned improved tolerance under a monochromatic forward model. The simulations results in section 4 showed that a polychromatic ptychography experiment likewise benefits from a curved illumination.

In essence, with no additional complexity in hardware requirements, ptychography experiments benefit from faster scanning (i.e, having a larger spot) and relaxed spectral bandwidth requirements, by simply defocusing the probe. We believe these finding have important bearings on the achievable throughput in photon-limited ptychography experiments, including inherently broadband tabletop high-harmonic generation [40] and x-ray sources [18].

**Funding.** Fraunhofer-Gesellschaft (Cluster of Excellence Advanced Photon Sources); Helmholtz Associa-

**Disclosures.** The authors declare no conflicts of interest.

**Data availability.** Data underlying the results presented in this paper are not publicly available at this time but may be obtained from the authors upon reasonable request.